\documentclass[%
 reprint,
 amsmath,amssymb,
 aps,
]{revtex4-2}

\usepackage{graphicx}
\usepackage{dcolumn}
\usepackage{bm}

\makeatletter

\newcommand{\Rmnum}[1]{\expandafter\@slowromancap\romannumeral #1@}
\makeatother

\def\<{\langle}
\def\>{\rangle}

\begin{document}

\title{Long Electron Spin Coherence Times of Atomic Hydrogen Trapped in Silsesquioxane Cages}

\author{George Mitrikas}
\email{g.mitrikas@inn.demokritos.gr}
\affiliation{Institute of Nanoscience and Nanotechnology, NCSR Demokritos, Aghia Paraskevi Attikis, 15310 Athens, Greece}


\date{\today}

\begin{abstract}
  Encapsulated atomic hydrogen in cube-shaped octa-silsesquioxane (POSS) cages of the Si$_8$O$_{12}$R$_8$ type (where R is an organic group) is the simplest alternative stable system to paramagnetic endohedral fullerenes (N@C$_{60}$ or P@C$_{60}$) that have been regarded as key elements of spin-based quantum technologies. Apart from common sources of decoherence like nuclear spin and spectral diffusion, all H@POSS species studied so far suffer from additional shortening of $T_2$ at low temperatures due to methyl group rotations. Here we eliminate this factor for the first time by studying the relaxation properties of the smallest methyl-free derivative of this family with R=H, namely H@T$_8$H$_8$. We suppress nuclear spin diffusion by applying dynamical decoupling methods and we measure electron spin coherence times $T_2$ up to 280 $\pm$ 76 $\mu$s at $T=90$ K. We observe a linear dependence of the decoherence rate $1/T_2$ on trapped hydrogen concentrations ranging between 9$\times 10^{14}$ cm$^{-3}$ and 5$\times 10^{15}$ cm$^{-3}$ which we attribute to the spin dephasing mechanism of instantaneous diffusion and a nonuniform spatial distribution of encapsulated H atoms.
\end{abstract}

\maketitle




%
Spin-based quantum computing is an active field of research exploring the use of spin particles as quantum bits (qubits). Electron and nuclear spins are particularly important in this context because they are natural quantum objects with relatively long coherence times that can be controlled using well-known magnetic resonance methods \cite{Morton_review2011,Mortonrev}. Achieving long coherence times is a significant challenge in this line of research and different molecular systems are continuously being evaluated as qubit candidates \cite{Wasielewski2020,reviewQIS2021}. Paramagnetic endohedral fullerenes (e.g. N@C$_{60}$ or P@C$_{60}$ with electron spin $S=3/2$) have received increased attention mainly because they provide a bottom-up route to large-scale quantum register fabrication and because they posses the longest electron spin coherence times of any molecular spin studied to date \cite{Harneit2017}. For P@C$_{60}$ Naydenov \cite{NaydenovPhD2006} reported a maximum $T_2$ value of 113 $\mu$s obtained with a two-pulse echo sequence (that was extended to 417 $\mu$s using dynamical decoupling methods) at $T=$10 K for a low spin concentration of 6.3$\times 10^{13}$ cm$^{-3}$, whereas, Brown and co-workers \cite{Brown2011} measured $T_2=$190 $\mu$s (that could be extrapolated to 300 $\mu$s in the limit of infinitively short refocusing pulses) at $T=$70 K for N@C$_{60}$ with a concentration of 2.5$\times 10^{15}$ cm$^{-3}$.

In 1994 Matsuda and co-workers \cite{Matsuda1994} discovered that upon $\gamma-$irradiation POSS cages can stably trap hydrogen atoms even at room temperature. Atomic hydrogen is the simplest paramagnetic atom with the electron spin $S=1/2$ coupled to the proton nuclear spin $I=1/2$ with a large hyperfine coupling constant of 1420.406 MHz. Therefore, the exceptional high stability of atomic hydrogen encapsulated in POSS (H@POSS) triggered several electron paramagnetic resonance (EPR) studies to compare their spin relaxation properties with those of endofullerens. Unlike C$_{60}$, which is virtually free from nuclear spin noise due to the low natural abundance (1.07 \%) of $^{13}$C, Si$_8$O$_{12}$R$_8$ cages constitute concentrated $^1$H nuclear spin systems, therefore, the electron spin coherence in H@POSS is dictated, as a rule, by nuclear spin diffusion \cite{Eaton2000,Eatonrev}. Indeed, early pulsed EPR works on alkyl-substituted POSS derivatives \cite{Dinse2001,Paech2006b} revealed a square exponential behavior of the Hahn echo decay at ambient temperature with $T_2$ of the order of 10 $\mu$s, in line with the above spin dephasing mechanism.

At temperatures below 200 K all studies published so far reported a shortening of $T_2$ to about 1 $\mu$s which was not reversible even at liquid helium temperatures. Using R groups with different rotational degrees of freedom we showed that this peculiarity, initially ascribed to changes in cage symmetry \cite{Dinse2001}, has its origin to the methyl rotation of organic groups \cite{Mitrikas2012,Mitrikas2020}. Moreover, our recent study with deuterated methyl groups provided strong evidence that the short $T_2$ values observed at very low temperatures could be assigned to quantum rotational tunneling \cite{Mitrikas2021}.

Herein we study for the first time the electron spin relaxation properties of H@Si$_8$O$_{12}$H$_8$, also known as H@T$_8$H$_8$, which is the smallest derivative of octa-silsesquioxanes. Interestingly, this species is the less studied among all H@POSS systems, presumably because the proximal proton nuclear spins of R and the larger delocalization of the spin wave function were assumed to contribute much more to decoherence compared to larger species. On the other hand, since H@T$_8$H$_8$ contains no CH$_3$ units it is an ideal system free from dynamic processes with short correlation times like the rotation of methyl groups. Moreover, nuclear spin diffusion could be efficiently eliminated since deuterium isotopic substitution is straightforward for this system \cite{Burgy1990,Calzaferri1992}.

A practical challenge in studying H@T$_8$H$_8$ is the appearance of strong free radical signals in the $g\approx2$ region upon $\gamma-$irradiation. Although these signals are spectroscopically well-separated from the EPR signal of atomic hydrogen, the relaxation properties of the latter can be affected significantly, especially when the concentration of free radicals is quite high. Unlike the majority of studied H@POSS, the free radical signals in H@T$_8$H$_8$ are not affected by the presence of radical scavengers and appear to be quite stable when exposed to air \cite{Stoesser1997}. To minimize the effects of unwanted free radicals we followed a different method for hydrogen encapsulation, namely electric discharge that has been proved to create less than one tenth of the radicals generated by $\gamma-$ray irradiation for the same resulting hydrogen encapsulation yield \cite{Harima2010} (see SI for details).

\begin{figure}
  \includegraphics{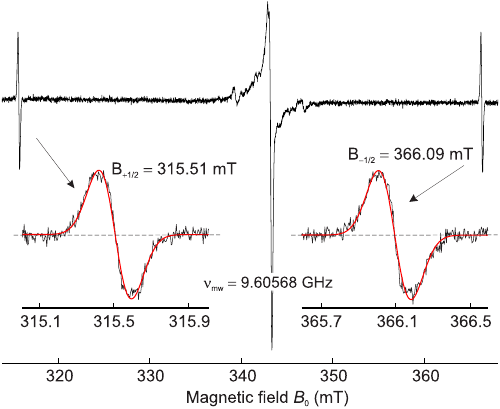}

  \caption{X-band room-temperature EPR spectrum of T$_8$H$_8$. The two insets show details of the H@T$_8$H$_8$ resonances corresponding to atomic hydrogen concentration of $C_{\mathrm{H}}=1.9\times 10^{15}$ cm$^{-3}$ (black traces) along with their best fitted simulations (red traces). For fitting parameters see text.} 
  \label{fig1}
\end{figure}

Fig. \ref{fig1} shows the room-temperature EPR signal of H@T$_8$H$_8$ corresponding to atomic hydrogen concentration $C_{\mathrm{H}}=1.9\times 10^{15}$ cm$^{-3}$. The obtained parameters $g=$2.00290(10), $A=$1410.5(2) MHz, and $\Delta B_{\text{pp}}=$174 $\mu$T were determined from numerical simulations using the isotropic spin Hamiltonian $\hat{\cal{H}}=g\beta_e B/hS_z-g_n\beta_n B_0/hI_z + A\mathbf{\hat{S}}\cdot\mathbf{\hat{I}}$ where $g$ and $g_n$ are the electron and nuclear $g$-factors, $\beta_e$ and $\beta_n$ are the Bohr and nuclear magnetons, $A$ is the isotropic hyperfine coupling of the encapsulated proton, $\Delta B_{\text{pp}}$ is the linewidth, and $B_0$ is the static magnetic field along \emph{z}-axis. These parameters are in good agreement with those obtained in previous studies \cite{Stoesser1997,Paech2006b} and verify the larger delocalization of the unpaired electron to the cage atoms for H@T$_8$H$_8$ as compared to all other POSS species \cite{Roduner2001}.

\begin{figure}
  \includegraphics{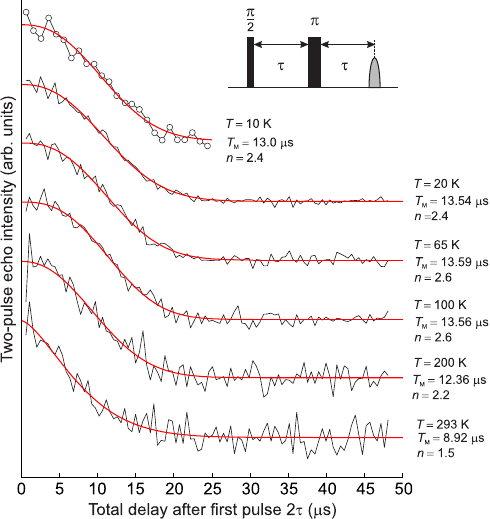}

  \caption{Two-pulse electron spin echo decays measured at six different temperatures as a function of 2$\tau$, and the superimposed stretched exponential fits using eq\ref{eqn1}. All traces were recorded at the observer position $B_0=$319.1 mT corresponding to the low-field EPR transition for the sample with $C_{\mathrm{H}}=1.9\times 10^{15}$ cm$^{-3}$.}
  \label{fig2}
\end{figure}

The transverse electron spin relaxation time, $T_2$, can be typically measured with the pulse sequence $\pi/2-\tau-\pi-\tau-$echo shown in Fig. \ref{fig2}. Decay traces measured at different temperatures show stretched-exponential behaviour that can be fitted with

\begin{equation}
  I(2\tau)=I_0\,\mathrm{exp}\left[- \left (\frac{2\tau}{T_{\mathrm{M}}}\right)^n\right],
  \label{eqn1}
\end{equation}
where $\tau$ is the interpulse delay, $n$ is a parameter determined by the mechanism of phase memory decay and the rate, $W$, of the dephasing process relative to $\tau$, and $T_{\mathrm{M}}$ is the so-called phase memory time encompassing $T_2$ and all other processes that cause electron spin dephasing \cite{Eaton2000}. The experimentally determined range of parameter $n$, $1.5\leq n\leq2.6$, implies a slow dynamic process with $W\tau \ll 1$.  For systems of low paramagnetic concentration and proton-containing ligands like the ones presented here, a very effective dephasing mechanism is the so-called nuclear spin diffusion \cite{Brown}. According to this, two neighbouring proton nuclear spins can undergo mutual spin flips with typical rates of $W/2\pi \sim$ 10 kHz, which in turn modulate the electron-nuclear dipolar interaction.

\begin{figure}
  \includegraphics{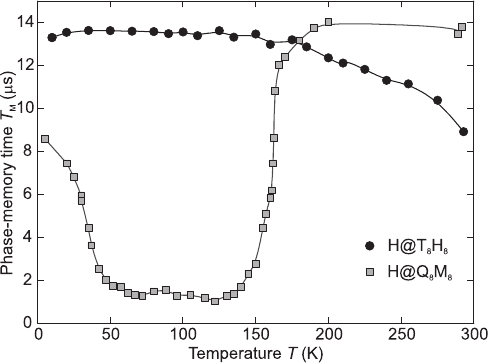}

  \caption{Temperature dependence of phase memory times $T_{\mathrm{M}}$ for H@T$_8$H$_8$ with $C_{\mathrm{H}}=1.9\times 10^{15}$ cm$^{-3}$ (R=H, circles) and H@Q$_8$M$_8$ (R=OSi(CH$_3$)$_3$, squares, modified with permission from \cite{Mitrikas2012} copyright © the Owner Societies 2020). Curves connect data points.}
  \label{fig3}
\end{figure}

The temperature dependence of $T_{\mathrm{M}}$ is shown in Fig. \ref{fig3}. Contrary to methyl-containing POSS derivatives (see for instance gray squares depicting previously published data for H@Q$_8$M$_8$), $T_{\mathrm{M}}$ becomes maximum around 150 K and remains constant in the temperature interval 10-150 K with a mean value of 13.4 $\mu$s. At room temperature, the obtained $T_{\mathrm{M}}=$8.9 $\mu$s is larger than the previously reported value of 3.8 $\mu$s, \cite{Paech2006b} but this difference could be ascribed to a possible larger concentration of paramagnetic centers in the previous case. The modest temperature dependence of $T_{\mathrm{M}}$ between 150 K and 293 K is assigned to the short spin-lattice relaxation times $T_1$ that range between 115 and 14 $\mu$s, respectively, and determine $T_{\mathrm{M}}$ in this temperature interval (see SI for details).

\begin{figure*}
  \includegraphics{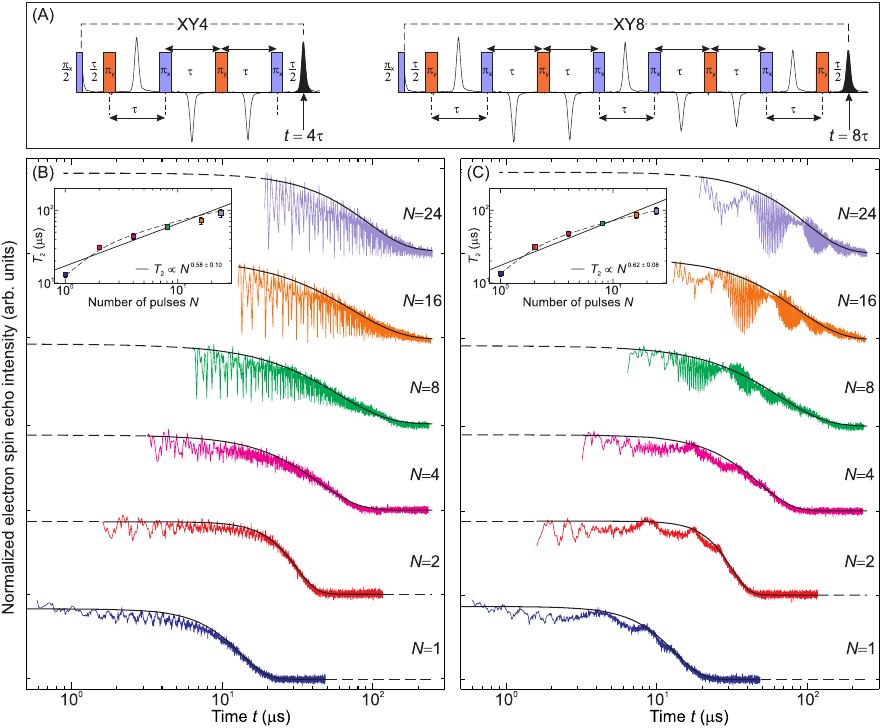}

  \caption{(A) XY4 and XY8 pulse sequences measuring the intensity of the last refocused echo (marked in black) as a function of the total evolution time $t$ after the $\pi/2$ pulse. (B) Electron spin coherence of H@T$_8$H$_8$ with $C_{\mathrm{H}}=1.9\times 10^{15}$ cm$^{-3}$ measured at $T=90$ K with DD sequences with different number of pulses at the low-field EPR transition ($B_0=$319.1 mT). $N=1$ corresponds to two-pulse echo decay; $N=2$ uses the sequence $(\pi/2)_x-\tau/2-(\pi)_y-\tau-(\pi)_x-\tau/2-\mathrm{echo}$; $N=4$ corresponds to XY4; $N=8, 16, 24$ correspond to XY8-1, XY8-2, and XY8-3, respectively. Black traces depict best fits with $I=I_0\cdot\mathrm{exp}(-t/T_2)^n$. Inset: Scaling of derived $T_2$ values with the number $N$ of DD pulses and the fitted curve $T_2\propto N^x$; dashed curve connects points as a guide to show the data trend. (C) Same as in (B) measured at the high-field EPR transition ($B_0=$369.9 mT)}
  \label{fig4}
\end{figure*}

An important aspect of these results is the absence of dynamic effects associated with methyl rotations that were previously observed in all H@POSS species. This paves the way for conducting experiments at much lower temperatures where $T_2$ is not any more limited by $T_1$ which exceeds 1 ms below 90 K. The suppression of nuclear spin diffusion and the investigation of additional underlying decoherence mechanisms can be best performed with dynamical decoupling methods comprising successive refocusing microwave (mw) pulses which are separated by time delays $\tau$ that are much shorter than the correlation time $\tau_c$ of the dephasing mechanism \cite{Mortonrev}. The Carr–Purcell–Meiboom–Gill (CPMG) sequence \cite{Carr_Purcell_1954,Meiboom_Gill_1958}, $(\pi/2)_x\{-\tau/2-(\pi)_y-\tau/2-\mathrm{echo}\}^N$, is a typical dynamical decoupling method that performs very well in nuclear magnetic resonance (NMR) spectroscopy \cite{Suter2013}. However, in EPR spectroscopy the unwanted stimulated echo, which appears as a consequence of partial excitation and non-ideal mw pulses, overlaps with the desired refocused primary echo \cite{Zaripov2021}. Since this stimulated echo decays with $T_1$, the CPMG sequence could erroneously result in longer than real $T_2$ values and, therefore, care has to be exercised when used. To ensure reliable $T_2$ measurements, we have used the more robust XY4 and XY8 pulse sequences shown in Fig. \ref{fig4}(A) that eliminate such unwanted signals (see SI for details) \cite{Maudsley1986,Gullion1990}.

Fig. \ref{fig4}(B) and Fig. \ref{fig4}(C) show the time evolution of electron spin coherence measured with dynamical decoupling sequences that use different number of pulses, $N$. Starting from the simple two-pulse sequence ($N=1$), all traces show stretched exponential decay as well as modulations originating from weak anisotropic hyperfine couplings between the unpaired electron and nearby magnetic nuclei like $^1$H and $^{29}$Si, the so-called electron spin echo envelope modulation (ESEEM) effect \cite{Schweiger2001,Doorslaer2016}. As we showed in our previous studies the application of dynamical decoupling methods in such spin systems can greatly enhance decoherence for specific values of $\tau$ \cite{Mitrikas2014} and the degree of this enhancement depends on the strength of hyperfine coupling and $N$ \cite{Mitrikas2015}. In terms of noise spectrum of the system under study, the hyperfine-coupled $^1$H and $^{29}$Si nuclear spins can be regarded as a source of high-frequency noise whose effect is enhanced upon application of large number of pulses $N$. Therefore, although dynamical decoupling suppresses nuclear spin diffusion (low-frequency noise), it may also enhance decoherence if the system bears such a source of high frequency noise.

To reduce the influence of these deep modulations on the determination of coherence times, we consider only the points of maximum echo intensity that define the envelopes of coherence decay curves. Data analysis shows that the maximum $T_2=100\pm10 $ $\mu$s is obtained for $N=24$ in both low- and high-field measurements. The insets of Fig. \ref{fig4} depict the scaling of $T_2$ with $N$ which - within experimental error - agrees well with a $T_2\propto N^{2/3}$ behaviour expected for a Lorentzian noise spectrum when $\tau_c \gg T_2$ \cite{Sousa2009}. Under this condition, $T_2$ is expected to increase with increasing $N$ with an upper limit of $T_2^{max}=2T_1$, whereas, for $\tau_c \ll T_2$ no improvement of $T_2$ with $N$ occurs. Our data show a saturation trend of $T_2$ for $N>8$, however, the reached value of 100 $\mu$s is much smaller than $2T_1=2$ ms at this temperature. Interestingly, this value matches the correlation time $\tau_c \sim$ 100 $\mu$s that corresponds to the proton nuclear spin flip-flop rate, $W/2\pi \sim$ 10 kHz. Consequently, we can assume that for proton nuclear spin diffusion no significant improvement of $T_2$ with $N>24$ should be expected for the system under study.

\begin{figure}
  \includegraphics{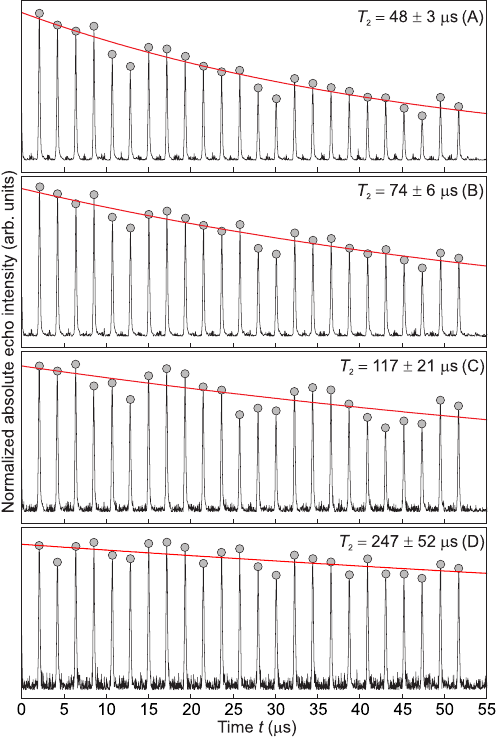}

  \caption{Time evolution of the spin magnetization under the application of the XY8-3 sequence with $\tau=2160$ ns for four H@T$_8$H$_8$ samples with different encapsulated hydrogen concentrations, $C_{\mathrm{H}}=4.9\times 10^{15}$ cm$^{-3}$ (A), $3.4\times 10^{15}$ cm$^{-3}$ (B), $1.9\times 10^{15}$ cm$^{-3}$ (C), and $9\times 10^{14}$ cm$^{-3}$ (D). Gray circles mark the refocused echo amplitudes; red traces depict their mono-exponential fits with $I=I_0\cdot\mathrm{exp}(-t/T_2)$. All measurements were performed at $T=90$ K at the observer position $B_0=$369.9 mT corresponding to the high-field EPR transition.}
  \label{fig5}
\end{figure}

Technical limitations of our spectrometer (maximum allowed number of mw pulses $N=30$ and maximum evolution time of 240 $\mu$s) do not allow for using additional mw pulses and test any possible improvement with $N$. On the other hand, since the correlation time of nuclear spin diffusion is $\tau_c \sim$ 100 $\mu$s, an effective dynamical decoupling sequence should use interpulse delays  $\tau\ll\tau_c $, i.e. shorter than 10 $\mu$s in order to suppress this dephasing mechanism. Therefore, instead of using sequences with varying $\tau$, we can set a constant $\tau$ value and measure the train of refocusing electron spin echoes occurring at times $t=\tau, 2\tau, ..., N\tau$ after the preparation $\pi/2$-pulse. With the proper choice of a short enough $\tau$ value, which at the same time corresponds to $^1$H and $^{29}$Si revivals of the spin-echo signals of Fig. \ref{fig4}, this setup ensures simultaneous suppression of both low- and high-frequency noise and allows for probing other dephasing mechanisms. Typical experiments using the XY8-3 sequence ($N=24$) with $\tau=2160$ ns are shown in Fig. \ref{fig5}.

The $T_2$ values obtained with the above dynamical decoupling method can be further analyzed based on the general formula of phase relaxation rate
\begin{equation}
  \frac{1}{T_2}=\frac{1}{T_{\mathrm{SD}}}+\frac{1}{T_{\mathrm{ID}}}.
  \label{eqn2}
\end{equation}
The first term, referred to as spectral diffusion, describes decoherence of the central spin (A spins) due to random fluctuations of dipole fields created by neighbour electron spins (B spins) that are not excited by the mw pulses. These fluctuations can either originate from spin-lattice relaxation of B spins ($T_1$-spectral diffusion) or mutual spin flips among them ($T_2$-spectral diffusion). For the first case, the contribution to phase relaxation rate is given by \cite{Schweiger2001}
  \begin{equation}
  \frac{1}{T_{\mathrm{SD}}}=\frac{1}{1.4}\sqrt{2.53\frac{\mu_0}{4\pi\hbar}g_Ag_B{\mu_B}^2\frac{C_\mathrm{B}}{{T_1}^{(\mathrm{B})}}},
  \label{eqn3}
\end{equation}
where ${T_1}^{(\mathrm{B})}$ is the spin-lattice relaxation of B spins and $C_\mathrm{B}$ is their concentration. To inspect if this type of spectral diffusion dominates our results, measurements at lower temperatures were performed. At $T=20$ K, where $T_1=550$ ms, Eq.\ref{eqn3} predicts a 23-fold increase of $T_{\mathrm{SD}}$ compared to $T=90$ K where $T_1=1$ ms. Our data show no sign of temperature dependence for $T_2$ (see SI for details) and therefore we conclude that $T_1$-spectral diffusion is not the dominant dephasing mechanism in our case.

The second term of Eq.\ref{eqn2} is the so-called instantaneous diffusion describing the static spread of the Larmor spin frequencies among excited dipole-coupled A spins which is imposed by the applied mw pulses. The contribution to phase relaxation rate is given by \cite{Schweiger2001}
\begin{equation}
  \frac{1}{T_{\mathrm{ID}}}=C_\mathrm{A}\frac{4\pi^2}{9\sqrt{3}}\frac{\mu_0}{4\pi\hbar}g_A^2{\mu_B}^2\mathrm{sin}^2\frac{\theta_2}{2}=C_\mathrm{A} k \mathrm{sin}^2\frac{\theta_2}{2},
  \label{eqn4}
\end{equation}
where $C_\mathrm{A}$ is the concentration of A spins, $k=8.2834\times10^{-13} \mathrm{cm}^3\mathrm{s}^{-1}$, and $\theta_2$ is the rotation angle of the refocusing pulse in the two-pulse sequence. A standard method to mitigate the effect of instantaneous diffusion is to measure two-pulse echo decays with small rotation angles $\theta_2$ and then extrapolate the data to the limit of infinitively short refocusing pulses in order to estimate the $T_2$ that is free from the instantaneous diffusion effect. Apparently, since our $T_{\mathrm{M}}$ values obtained with the two-pulse sequence are completely masked by nuclear spin diffusion, this methodology can not be applied here. However, as $T_{\mathrm{ID}}$ depends on $C_\mathrm{A}$, one can probe the contribution of instantaneous diffusion by comparing the $T_2$ values obtained with dynamical decoupling experiment on samples with different encapsulated hydrogen concentrations $C_\mathrm{H}$.

Fig. \ref{fig5} shows the data measured with the XY8-3 sequence with $\tau=2160$ ns for four samples with different $C_\mathrm{H}$. Clearly, as the encapsulated hydrogen concentration is reduced, the intensity of the refocused echoes is retained for longer evolution times. For the most diluted sample with $C_\mathrm{H}=9\times 10^{14}$ cm$^{-3}$ dynamical decoupling experiments obtain $T_2=247$ $\pm$ 52 $\mu$s which is the longest electron spin coherence time ever measured for the H@POSS system. Again, the limited number of available mw pulses does not allow for observing the full echo decay and thus determining $T_2$ with higher accuracy. Covering the whole necessary time window with $N=24$ pulses requires sequences with $\tau \geq$ 10 $\mu$s which is, however, of no use because nuclear spin diffusion dominates electron spin dephasing in this case. We anticipate that elimination of nuclear spin diffusion by deuterium isotopic substitution will make such scheme possible in our future studies.

\begin{figure}
  \includegraphics{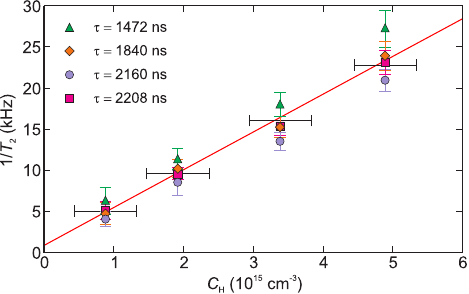}

  \caption{Phase relaxation rates versus encapsulated hydrogen concentration $C_\mathrm{H}$ for data acquired with four different $\tau$. The straight line is the linear fit with eq\ref{eqn5} giving $\alpha_\mathrm{M}=11.1\pm0.7$ and $b_0=871\pm934$ Hz.}
  \label{fig6}
\end{figure}

The determined phase relaxation rates 1/$T_2$ correlate very well with $C_\mathrm{H}$ as can be seen in Fig. \ref{fig6} where data from measurements with four different $\tau$ values are collected. The apparent linear dependence suggests either instantaneous or $T_2$-type spectral diffusion. It should be noted that, although the dynamical decoupling methods used here can efficiently suppress spectral diffusion mechanisms with correlation times $\tau_c\geq10$ $\mu$s, they can not refocus interactions between identical spins, so they are completely ineffective in suppressing instantaneous diffusion. Therefore, we can assume that the $C_\mathrm{H}$ dependence of 1/$T_2$ is virtually governed by the mechanism of instantaneous diffusion.

To further test this assumption, we model our data with a modified version of eq\ref{eqn4}
\begin{equation}
  \frac{1}{T_2}=b_0+\alpha_\mathrm{M} \frac{C_\mathrm{H}}{2}k,
  \label{eqn5}
\end{equation}
where $b_0$ is a constant and $C_\mathrm{A}$ has been replaced by $C_\mathrm{H}/2$ because since the two EPR transitions are well separated the measurement on each one of them involves only half of the encapsulated hydrogen atoms. $\alpha_\mathrm{M}$ is a scaling factor accounting for a possible deviation of the local spin concentration $C_\mathrm{loc}=\alpha_\mathrm{M}C_\mathrm{H}$ from $C_\mathrm{H}$, the average spin concentration of the encapsulated hydrogen atoms as determined from continuous wave EPR spectroscopy. The linear fit of data with eq\ref{eqn5} gives $\alpha_\mathrm{M}=11.1\pm0.7$, i.e. $C_\mathrm{loc}\approx 11$ $C_\mathrm{H}$, implying a nonuniform spatial distribution of paramagnetic centers ($C_\mathrm{loc} \geq C_\mathrm{H}$), which is a well-known result of track effects in irradiated solids \cite{Salikhov}. We have previously observed similar differences between $C_\mathrm{loc}$ and $C_\mathrm{av}$ for low-dose $\gamma-$irradiated POSS cages \cite{Mitrikas2021}. Interestingly, the method of electric discharge used in the present work favors the trapping of H atoms mainly on the surfaces of the molecular crystals \cite{Harima2010}, a fact that can adequately justify the aforementioned nonuniform spatial distribution of encapsulated H atoms.

In conclusion, we have managed for the first time to measure long electron spin coherence times $T_2$ up to 280 $\pm$ 76 $\mu s$ at $T=90$ K for the smallest H@POSS molecule, namely H@T$_8$H$_8$. The essence of this unprecedented level of improvement lies in the lack of methyl rotations that were previously present in all studied H@POSS species acting as the dominant dephasing mechanism especially at low temperatures. Our results showed that instantaneous diffusion is the only limiting decoherence mechanism for H@T$_8$H$_8$ as all other important mechanisms could be suppressed by dynamical decoupling. For real applications it may also be necessary to physically reduce the sources of these mechanisms in order to best exploit the long coherence times of this species and simplify the pulse sequences for building efficient quantum gates. For the case of nuclear spin diffusion this can be easily tackled with deuterium isotopic substitution. Further increase of $T_2$ depends on the ability to achieve a uniform spatial distribution of H atoms and to control their concentration. Although this may require some progress in POSS chemistry to be made, our study showed for the first time the potential of H@T$_8$H$_8$ in spin-based quantum technologies as it can equally compete endohedral fullerenes in terms of coherence times.

%
%





\bibliography{jpcl}

\end{document}


\begin{center}
\textbf{Electronic Supplementary Information}
\\
\hfill
\\
for the manuscript
\\
\hfill
\\
\textbf{\Large Long Electron Spin Coherence Times of Atomic Hydrogen Trapped in Silsesquioxane Cages}
\\
\hfill
\\
by
\\
\hfill
\\
\large {George Mitrikas}
\\
\hfill
\\
\normalsize
\emph{Institute of Nanoscience and Nanotechnology, NCSR Demokritos, 15310 Athens, Greece}
\\
\hfill
\\
E-mail: g.mitrikas@inn.demokritos.gr
\end{center}

\section{\label{sec:level6} Sample Preparation}

\textbf{POSS synthesis}: The hydridospherosiloxane (HSiO$_{3/2}$)$_8$, also abbreviated as T$_8$H$_8$, was prepared either with the original method of Agaskar \cite{Agaskar1991} or with the new synthesis of Tsukada and co-workers \cite{Tsukada2021}.

\noindent
\textbf{Hydrogen encapsulation}: Encapsulation of atomic hydrogen was first performed with $\gamma$-irradiation using a $^{60}$Co source (dose rate of about 5 kGy/day). The accumulated dose was measured using Red Perspex Dosimeters, Type 4034 AD. Typically 60 mg of POSS powder were placed in a sealed vial and irradiated for 12 days resulting in a total dose of 60 kGy. Comparison of the cw EPR spectrum (in the presence of O$_2$, unsaturated conditions) with a standard sample gave an estimate of $C_{\mathrm{H}}=8.6\times$10$^{15}$ spins per cm$^{3}$ for the electron spin concentration.

\begin{figure}[!t]
\centering
  \includegraphics{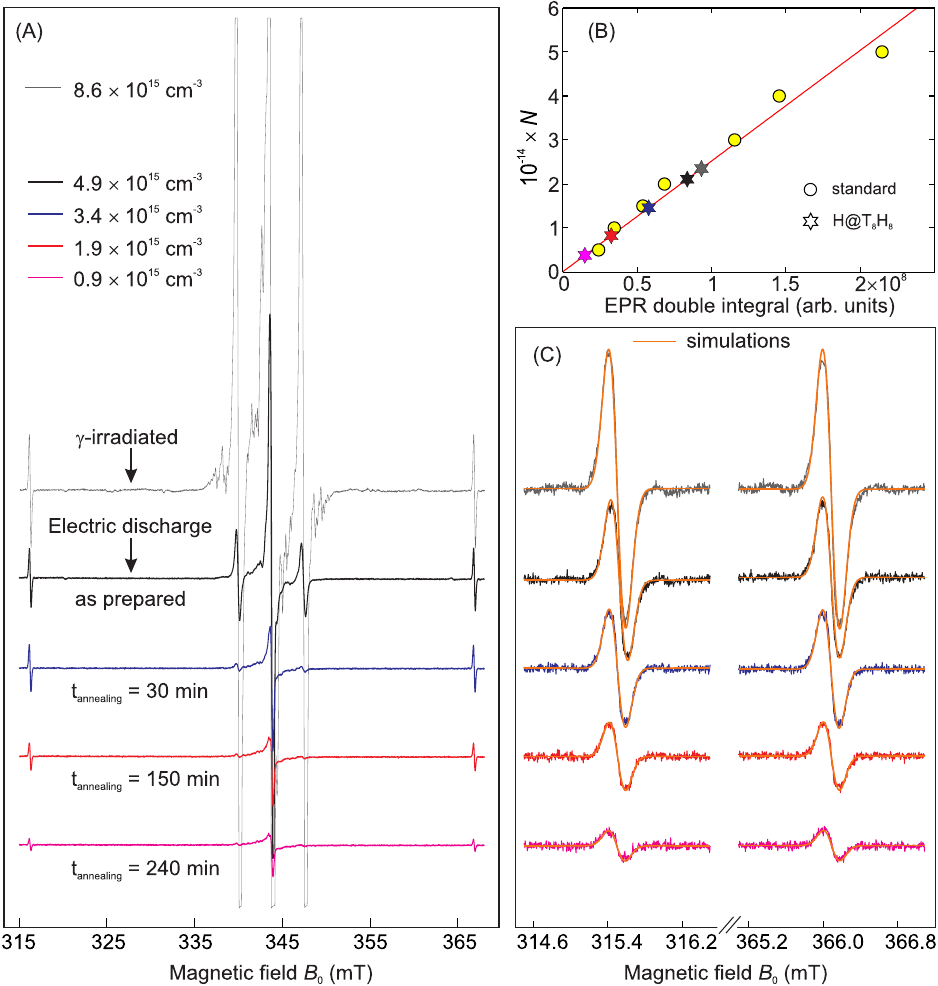}
\caption{(A) X-band room-temperature EPR powder spectra of the $\gamma$-irradiated sample ($m=53.7$ mg) and the four samples ($m=85$ mg) prepared with electric discharge. The signal intensity is normalized to the mass $m$ of the samples. Experimental conditions: microwave frequency, 9.621 GHz; microwave power attenuation, 40 dB; modulation amplitude, 0.15 mT; modulation frequency, 100 kHz; number of accumulated spectra, 2. (B) Calibration curve (red line) obtained from the standard samples denoted by circles. Star symbols depict data from H@T$_8$H$_8$ samples. (C) Details of the H@T$_8$H$_8$ resonance lines together with the simulated spectra (orange lines). Experimental conditions: microwave frequency, 9.621 GHz; microwave power attenuation, 40 dB; modulation amplitude, 25 $\mu$T; modulation frequency, 100 kHz; number of accumulated spectra, 10.}
\label{fig:S1}       
\end{figure}

To avoid the strong free radical signals in the $g\approx2$ region (gray trace in FIG.\ref{fig:S1}(A)), we followed the electric discharge method \cite{Harima2010} using a Tesla coil of maximum voltage 30 kV. About 100 mg of POSS powder were placed in a Schott test tube filled with air of 1 mbar pressure. The bottom of the test tube was placed close to a home-build Tesla coil and the glow discharge was applied for 1 min in three steps of 20 s with some waiting time between them to let the sample cool down. After this step, some powder that stacked on the glass wall was scratched from the tube and stirred with the rest of the material using a glass rod. The process was repeated three times with a total irradiation time of 3 min. 85 mg of the material were transferred in an EPR tube and the cw EPR spectrum was recorded at room temperature (as prepared sample, $C_{\mathrm{H}}=4.9\times 10^{15}$ cm$^{-3}$).

After completion of pulsed EPR measurements on this sample, the concentration of encapsulated hydrogen atoms $C_{\mathrm{H}}$ was reduced by annealing the powder for some time at $T=90^{\circ}$ C. This was done by placing the EPR tube in a furnace and following the intensity of the room-temperature H EPR signal every 10 min. With this process the $C_{\mathrm{H}}$ could be adjusted in a rather controlled way because the mass of the sample was constant. When the desired $C_{\mathrm{H}}$ was achieved, the sample was measured with pulsed EPR. Three more successive annealing cycles were repeated this way, reducing the concentration of the sample to $3.4\times 10^{15}$ cm$^{-3}$, $1.9\times 10^{15}$ cm$^{-3}$, and $0.9\times 10^{15}$ cm$^{-3}$ respectively. Their room-temperature EPR spectra are shown in FIG.\ref{fig:S1}(A).

\noindent
\textbf{Determination of $C_{\mathrm{H}}$}: The encapsulated H concentration of samples was determined based on a calibration curve created with the standard signal of Picein 80 \cite{picein80} having a known concentration of 1.97 $\pm$ 0.10 $\times 10^{15}$ spins/mg. To create samples of low amount of spins that are comparable with those of H@T$_8$H$_8$, 10 mg of Picein 80 were dissolved in 10 ml of benzene and seven EPR tubes were filled with 25, 50, 75, 100, 150, 200 and 250 $\mu$l of this solution, containing $N=0.5\times 10^{14}$, $1\times 10^{14}$, $1.5\times 10^{14}$, $2\times 10^{14}$, $3\times 10^{14}$, $4\times 10^{14}$, and $5\times 10^{14}$  spins, respectively. After evaporation of benzene the seven samples were measured and the calibration curve of FIG.\ref{fig:S1}(B) was obtained. The number of spins in the five H@T$_8$H$_8$ samples were estimated by comparing the double integral of each cw EPR spectrum with the calibration curve. To reduce baseline effects, especially for low concentration samples, the double integral of the simulated spectra shown in FIG.\ref{fig:S1}(C) were used. Finally, the concentration was estimated from $C_{\mathrm{H}}=Nd/m$, where $N$ is the number of spins, $d=1.97$ g/cm$^{-3}$ is the density of T$_8$H$_8$, and $m$ is the mass of the sample.

\section{\label{sec:level1} Spectroscopy}

EPR measurements at the X-band were carried out on a Bruker ESP 380E spectrometer equipped with an EN 4118X-MD4 Bruker resonator (both cw and pulse mode). Low temperature measurements were performed with a helium cryostat from Oxford Inc (at temperatures between 10 and 300 K). The microwave frequency was measured with a HP 5350B microwave frequency counter. The temperature was stabilized with an Oxford ITC4 temperature controller within  $\pm$0.1 K. The magnetic field was calibrated using a DPPH standard. The relaxation measurements were performed at both low- and high-field EPR transitions. The repetition rate was properly adjusted in every measurement in order to avoid saturation.

The electron spin–lattice relaxation times $T_1$ were measured by inversion recovery with the pulse sequence $\pi-t-\pi/2-\tau-\pi-\tau-$echo. The lengths of the mw $\pi/2$ and $\pi$ pulses were 16 and 32 ns, respectively, and the interpulse delay $\tau=$ 600 ns. For each trace, 100 data points were collected with an appropriate time increment to ensure complete magnetization recovery. The phase-memory times $T_\text{M}$ were measured by the two-pulse echo decay sequence, $\pi/2-t-\pi-t-$echo, with $t$ varying.

At temperatures lower than 20 K, where $T_1$ exceeds the maximum instrumental shot repetition time of 2 s, the electron spin echo was recorded with a HP Infinium 54810A oscilloscope which allowed the acquisition of the entire time trace in a single shot. The corresponding $T_1$ values were obtained with the saturation method: after saturation using a fast repetition rate, the microwave irradiation was interrupted and the system allowed to relax for time $t$ which was recorded with a chronometer. The recovered magnetization was measured with a two-pulse echo in a single-shot experiment recorded with the oscilloscope.

For dynamical decoupling experiments with constant $\tau$ values (Fig.5 of main paper), the train of refocused echoes was recorded with a HP Infinium 54810A oscilloscope after averaging out 1024 measurements acquired with a shot repetition time of 10 ms at $T=90$ K (this was properly increased for measurements at lower temperatures) and a time resolution of 20 ns. To remove pulse signals and baseline artifacts, we additionally measured the same traces in an out-of-resonance position were all echoes were zero, and the two traces were subtracted.

The data were processed with the program MATLAB 7.0 (The MathWorks, Natick, MA). $T_1$ relaxation times were determined
by fitting the time traces with single exponential functions. For obtaining $T_\text{M}$ a stretched exponential function was used. cw-EPR spectra were simulated with the EasySpin package \cite{Stoll2006}.

\section{\label{sec:level6} Spin-Lattice Relaxation Times $T_1$}
\begin{figure}[!h]
\centering
  \includegraphics{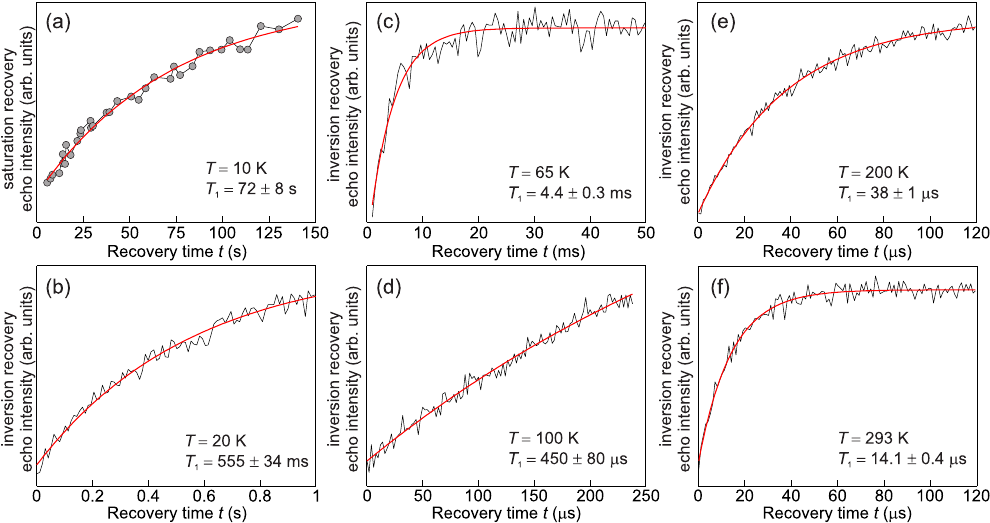}
\caption{Saturation (a) and inversion (b–f) recovery data at six different temperatures and $T_1$ values obtained from the superimposed exponential fits.}
\label{fig:S2}       
\end{figure}

FIG.\ref{fig:S2} illustrates spin-lattice relaxation data at the same temperatures corresponding to $T_\text{M}$ measurements shown in Fig.2 of main paper.

\section{\label{sec:level6} XY8 vs CPMG}
\begin{figure}[!h]
\centering
  \includegraphics{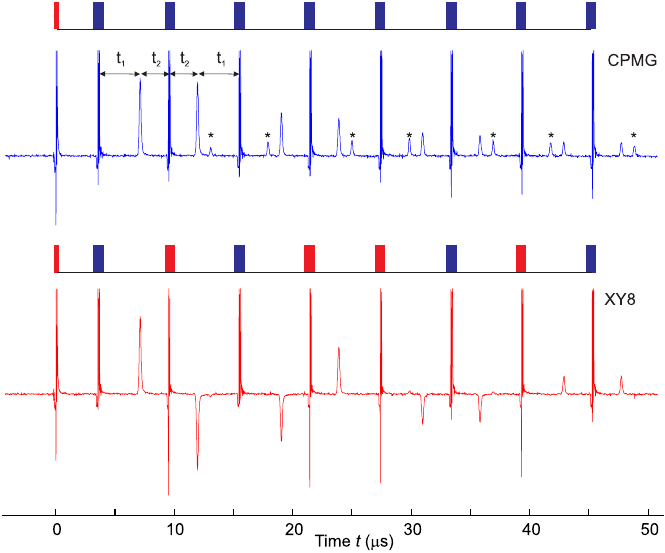}
\caption{Comparison of CPMG (top) and XY8 sequences for $t_1=3536$ ns and $t_2=2432$ ns. Asterisks denote the positions of stimulated echoes appearing in CPMG.}
\label{fig:S3}       
\end{figure}

In order to verify that the XY8 pulse sequence eliminates the unwanted stimulated echoes that are present in CPMG, we used $t_1\neq t_2$ that separates refocused from stimulated echoes. FIG.\ref{fig:S3} shows that for CPMG the stimulated echoes have significant intensity, whereas for XY8 they are virtually absent.

\section{\label{sec:level6} Temperature dependence of $T_2$}
\begin{figure}[!h]
\centering
  \includegraphics{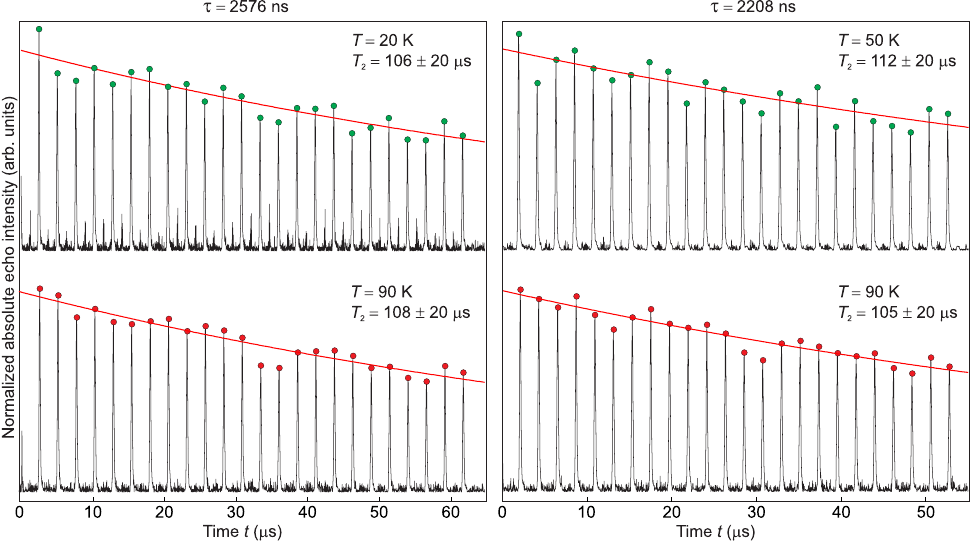}
\caption{Time evolution of the spin magnetization under the application of the XY8-3 sequence for the sample with $C_{\mathrm{H}}=1.9\times 10^{15}$ cm$^{-3}$.}
\label{fig:S4}       
\end{figure}

To inspect whether the $T_2$ values obtained with dynamical decoupling are determined by $T_1$-spectral diffusion, measurements at lower than 90 K were performed. FIG.\ref{fig:S4} shows that XY8-3 data acquired at 20 K and 50 K are virtually temperature independent.

\bibliography{jpcl}